\documentclass[ALICE,manyauthors]{cernphprep}

\usepackage[comma,square,numbers,sort&compress]{natbib}
\usepackage{hyperref}
\usepackage{lineno}
\usepackage{graphicx,subfigure}

\def\snn{\mbox{$\sqrt{s_{_{\rm NN}}}$}}
\def\pt{p_{\rm{T}}}

\begin{document}%

\begin{titlepage}
\PHyear{2017}
\PHnumber{103}      
\PHdate{10 May}  
%

\title{Linear and non-linear flow mode in Pb--Pb collisions at $\sqrt{s_{\rm NN}} =$ 2.76 TeV }
\ShortTitle{Linear and non-linear flow modes in Pb--Pb at 2.76 TeV}   

\Collaboration{ALICE Collaboration\thanks{See Appendix~\ref{app:collab} for the list of collaboration members}}
\ShortAuthor{ALICE Collaboration} 

\begin{abstract}


The second and the third order anisotropic flow, $V_{2}$ and $V_3$, are mostly determined by the corresponding initial spatial anisotropy coefficients, $\varepsilon_{2}$ and $\varepsilon_{3}$, in the initial density distribution. In addition to their dependence on the same order initial anisotropy coefficient, higher order anisotropic flow, $V_n$ ($n > 3$), can also have a significant contribution from lower order initial anisotropy coefficients, which leads to mode-coupling effects. In this Letter we investigate the linear and non-linear modes in higher order anisotropic flow $V_n$ for $n=4$, 5, 6 with the ALICE detector at the Large Hadron Collider. The measurements are done for particles in the pseudorapidity range $|\eta| < 0.8$ and the transverse momentum range $0.2 < p_{\rm T} < 5.0$ GeV/$c$ as a function of collision centrality. The results are compared with theoretical calculations and provide important constraints on the initial conditions, including initial spatial geometry and its fluctuations, as well as the ratio of the shear viscosity to entropy density of the produced system.


\end{abstract}
\end{titlepage}
\setcounter{page}{2}


\section{Introduction}
\label{sec:intro}

The primary goal of the ultra-relativistic heavy-ion collision programme at the Large Hadron Collider (LHC) is to study the properties of the Quark-Gluon Plasma (QGP), a novel state of strongly interacting matter that is proposed to exist at high temperatures and energy densities~\cite{Shuryak:1978ij,Shuryak:1980tp}. Studies of azimuthal correlations of produced particles have contributed significantly to the characterisation of the matter created in heavy-ion collisions~\cite{Ollitrault:1992bk,Voloshin:2008dg}. Anisotropic flow, which quantifies the anisotropy of the momentum distribution of final state particles, is sensitive to the event-by-event fluctuating initial geometry of the overlap region, together with the transport properties and equation of state of the system~\cite{Voloshin:2008dg, Heinz:2013th,Pratt:2015zsa,Song:2017wtw}. The successful description of anisotropic flow results by hydrodynamic calculations suggests that the created medium behaves as a nearly perfect fluid~\cite{Heinz:2013th,Voloshin:2008dg} with a shear viscosity to entropy density ratio, $\eta/s$, close to a conjectured lower bound $1/4\pi$~\cite{Kovtun:2004de}. Anisotropic flow is characterised using a Fourier decomposition of the particle azimuthal distribution in the plane transverse to the beam direction~\cite{Voloshin:1994mz, Poskanzer:1998yz}: 
\begin{equation}
\frac{dN}{d\varphi} \propto 1+2\sum_{n=1}^{\infty}v_{n} \cos[n(\varphi - \Psi_{n})],
\label{eq:FourierExp}
\end{equation}
where $N$ is the number of produced particles, $\varphi$ is the azimuthal angle of the particle and $\Psi_{n}$ is the $n^{th}$ order flow symmetry plane. The $n^{th}$ order (complex) anisotropic flow $V_{n}$ is defined as: $V_{n} \equiv v_{n} \, e^{in\Psi_{n}}$, where $v_{n}  = |V_{n}|$ is the flow coefficient, and $\Psi_{n}$ represents the azimuth of $V_{n}$ in momentum space. For non-central heavy-ion collisions, the dominant flow coefficient is $v_2$, referred to as elliptic flow. Non-vanishing values of higher flow coefficients $v_3$--$v_6$ at the LHC are ascribed primarily to the response of the produced QGP to fluctuations of the initial energy density profile of the colliding nucleons~\cite{Alver:2010gr, ALICE:2011ab, ATLAS:2012at, Chatrchyan:2013kba,Adam:2016izf}.

The standard (moment-defined) initial anisotropy coefficients $\varepsilon_n$ together with their corresponding initial symmetry planes (also called participant planes) $\Phi_{n}$ can be calculated from the transverse positions $(r, \phi)$ of the participating nucleons
\begin{equation}
\label{eq:eccCalculation}
\varepsilon_{n}e^{in\Phi_{n}} \equiv - \frac{\left< r^{n} e^{in\phi} \right>}{{\left< r^{n}  \right>}} \, \, \, \, ({\rm for} ~~n>1),
\end{equation}
where $\left< \, \right>$ denotes the average over the transverse position of all participating nucleons, $\phi$ is azimuthal angle, and $n$ is the order of the coefficient~\cite{Alver:2010gr,Alver:2010dn}. It has been shown in~\cite{Qiu:2011hf,Niemi:2012aj} that $V_{2}$ and $V_{3}$ are mostly determined with the same order initial spatial anisotropy coefficients $\varepsilon_2$ and $\varepsilon_{3}$, respectively. Considering that $\eta/s$ reduces the hydrodynamic response of $v_n$ to $\varepsilon_{n}$, it was proposed in~\cite{Song:2010mg, Niemi:2012aj,Gardim:2014tya, Fu:2015wba} that $v_{n} / \varepsilon_{n}$ (for $n = 2, 3$) could be a direct probe to quantitatively constrain the $\eta/s$ of the QGP in hydrodynamic calculations. However, $\varepsilon_n$ cannot be determined experimentally. Instead, they are obtained from various theoretical models, resulting in large uncertainties in the estimated $\eta/s$ derived indirectly from $v_{2}$ and $v_{3}$ measurements~\cite{Song:2010mg, Qiu:2011hf}. On the other hand, higher order anisotropic flow $V_{n}$ with $n > 3$ probe smaller spatial scales and thus are more sensitive to $\eta/s$ than $V_{2}$ and $V_{3}$ due to more pronounced viscous corrections~\cite{Alver:2010dn,Teaney:2012ke}. Thus, the study of the full set of flow coefficients is expected to constrain both $\varepsilon_{n}$ and $\eta/s$ simultaneously. However, it was realised later that $V_{n}$ with $n>3$ is not linearly correlated with the corresponding $\varepsilon_{n}$~\cite{Alver:2010dn,Gardim:2011xv,Teaney:2012ke}, which makes the extraction of $\eta/s$ from measurements of higher order flow coefficients less straightforward. In addition to the study of flow coefficients, the results of correlations between different order anisotropic flow angles and amplitudes shed light on both the early stage dynamics and the transport properties of the created QGP~\cite{Bilandzic:2013kga, Bhalerao:2014xra, Aad:2015lwa, ALICE:2016kpq, Zhou:2016eiz, Qiu:2012uy,Teaney:2013dta, Niemi:2015qia, Zhou:2015eya}. In particular, the characteristic pattern of flow symmetry plane correlations (also known as angular correlations of flow-vectors) observed in experiments is reproduced quantitatively by theoretical calculations~\cite{Qiu:2012uy,Teaney:2013dta, Niemi:2015qia,Aad:2014fla, Zhou:2015eya}. However, the correlations between flow coefficients (also known as amplitude correlations of flow-vectors), investigated using symmetric cumulants, provide stricter constraints on initial conditions and $\eta/s$ than the individual $v_n$ measurements~\cite{Bilandzic:2013kga, Bhalerao:2014xra, Niemi:2015qia, Aad:2015lwa, ALICE:2016kpq,Zhou:2016eiz, Zhou:2015eya}. It is a challenge for current theoretical models to provide quantitative descriptions of the correlations between different order flow coefficients. 

As discussed above, it is known that the lower order anisotropic flow $V_{n}$ ($n=$ 2, 3) is largely determined by a linear response of the system to the corresponding $\varepsilon_{n}$ (except in peripheral collisions). Higher order anisotropic flow $V_{n}$ with $n>$ 3 have contributions not only from the linear response of the system to $\varepsilon_{n}$, but also contributions proportional to the product of $\varepsilon_{2}$ and/or $\varepsilon_{3}$. These contributions are usually called non-linear response~\cite{Bhalerao:2014xra,Yan:2015jma} in higher order anisotropic flow. For a single event, $V_{n}$ with $n=$ 4, 5 and 6 can be decomposed into the so-called linear and the non-linear contributions, according to
\begin{eqnarray}
V_{4} &=&  V_{4}^{\rm NL} + V_{4}^{\rm L} = \chi_{4,22} (V_{2})^2 +  V_{4}^{\rm L} , \label{eq:V4}\\
V_{5} &=&  V_{5}^{\rm NL} + V_{5}^{\rm L} = \chi_{5,32} V_{2} \, V_{3} + V_{5}^{\rm L} , \label{eq:V5}\\
V_{6} &=&  V_{6}^{\rm NL} + V_{6}^{\rm L} =  \chi_{6,222} (V_{2})^3 + \chi_{6,33}(V_{3})^2 + \chi_{6,42} V_{2} V_{4}^{\rm L} + V_{6}^{\rm L}. \label{eq:V6}
\end{eqnarray}
Here $\chi_{n,mk}$ is a new observable called the non-linear mode coefficient~\cite{Yan:2015jma} and $V_{n}^{\rm NL}$ represents the non-linear mode which has contributions from modes with lower order anisotropy coefficients. The $V_{n}^{\rm L}$ term represents the linear mode, which was na\"{\i}vely expected from the linear response of the system to the same order $\varepsilon_{n}$. However, a recent hydrodynamic calculation showed that $V_{n}^{\rm L}$ is not driven by the linear response to the standardly moment-defined $\varepsilon_{n}$ introduced in Eq.~(\ref{eq:eccCalculation}), but the corresponding cumulant-defined anisotropy coefficient $\varepsilon_{n}^{'}$~\cite{Teaney:2013dta,Qian:2017ier}. For example, $V_{4}^{\rm L}$ is expected to be driven by the 4$^{th}$-order cumulant-defined anisotropy coefficient and its corresponding initial symmetry plane which can be calculated as
\begin{equation}
\label{eq:eccpCalculation}
\varepsilon_{4}^{'}e^{i4\Phi_{4}^{'}} \equiv - \frac{\left< z^{4} \right> - 3 \left< z^{2} \right>^{2}}{\left< r^{4} \right>} = \varepsilon_{4}e^{i4\Phi_{4}} + \frac{3 \left< r^{2} \right>^{2}}{\left< r^{4} \right>} \varepsilon_{2}^{2} e^{i4\Phi_{2}},
\end{equation}
where $z = r e^{i\phi}$. The calculations for other order anisotropy coefficients and their corresponding initial symmetry planes can be found in~\cite{Teaney:2013dta,Qian:2017ier}.
If the non-linear and linear modes of higher order anisotropic flow, $V_{n}^{\rm NL}$ and $V_{n}^{\rm L}$, are uncorrelated (e.g. $V_{n}^{\rm L}$ is perpendicular to $V_{n}^{\rm NL}$), they can be isolated. One of the proposed approaches to validate the assumption that $V_{n}^{\rm NL}$ and $V_{n}^{\rm L}$ are uncorrelated is testing the following relations~\cite{Bhalerao:2014xra}:
\begin{eqnarray}
\frac{ \left< V_{4} \, (V_{2}^{*})^{2} \, v_{2}^{~2} \right>} {  \left< V_{4}  \, (V_{2}^{*})^{2} \right>  \, \left<v_{2}^{~2} \right> } =  \frac{ \left< v_{2}^{~6} \right>} {  \left< v_{2}^{~4} \right>  \, \left<v_{2}^{~2} \right> } ,\label{eq:equality1} \\
\frac{ \left< V_{5} \, V_{3}^{*} \,  V_{2}^{*} \, v_{2}^{~2}  \right>} { \left< V_{5} \, V_{3}^{*} \,  V_{2}^{*}  \right> \, \left< v_{2}^{~2}\right>} =  \frac{ \left< v_{2}^{~4} \, v_{3}^{~2} \right>} {  \left< v_{2}^{~2} \, v_{3}^{~2} \right>  \, \left<v_{2}^{~2} \right> }.
\label{eq:equality2}
\end{eqnarray}

If the above relations are valid, one could combine the analyses of higher order anisotropic flow with respect to their corresponding symmetry planes and to the planes of lower order anisotropic flow $V_{2}$ or $V_{3}$ to eliminate the uncertainty from initial state assumptions and extract $\eta/s$ with better precision~\cite{Yan:2015jma}.

In this Letter, the linear and non-linear modes in higher order anisotropic flow generation are studied in Pb--Pb collisions at $\snn=$ 2.76 TeV with the ALICE detector. The main observables are introduced in Section~\ref{sec:probe} and the experimental setup is described in Section~\ref{sec:setup}. Section~\ref{sec:sys} presents the study of the systematic uncertainties of the above mentioned observables. The results and their discussion are provided in Section~\ref{sec:Results}. Section~\ref{sec:summary} contains the summary and conclusions.

\section{Observables and analysis methods}
\label{sec:probe}

Ideally, the flow coefficient $v_{n}$ can be measured via the azimuthal correlations of emitted particles with respect to the symmetry plane $\Psi_{n}$ as $v_{n} = \left < \cos n(\varphi - \Psi_{n}) \right>$. Since $\Psi_n$ is unknown experimentally, the simplest approach to obtain $v_{n}$ is using 2-particle correlations:
\begin{equation}
v_{n}\{2\} = \langle \langle \cos n( \varphi_{1} - \varphi_{2}) \rangle \rangle^{1/2} =  \left< v_{n}^{2} \right>^{1/2}.
\label{eq:vn2}
\end{equation}
Here $\langle \langle \, \rangle \rangle$ denotes the average over all particles in a single event and then an average over all events, $\left< \right>$ indicates the event average of over all events, and $\varphi_{i}$ represents the azimuthal angle of the $i$-th particle. The analysed events are divided into two sub-events A and B, separated by a pseudorapidity gap, to suppress non-flow effects. The latter are the azimuthal correlations not associated to the common symmetry plane $\Psi_{n}$, such as jets and resonance decays. Thus, we modify Eq.~(\ref{eq:vn2}) to
\begin{equation}
v_{n}\{2\} = \langle  \langle \cos (n \varphi_{1}^{A} - n\varphi_{2}^{B}) \rangle \rangle^{1/2}  =  \left< v_{n}^{2} \right>^{1/2}.
\label{eq:cnnn}
\end{equation}
Here $\varphi_{1}^{A}$ and $\varphi_{2}^{B}$ are selected from subevent A and B, respectively.

Before introducing observables related to the linear and non-linear modes in higher order anisotropic flow, it is crucial to verify whether Eqs.~(\ref{eq:equality1}-\ref{eq:equality2}) are applicable.
The left and right hand sides of Eq.~(\ref{eq:equality1}) are obtained by constructing suitable multi-particle correlations~\cite{Yan:2015jma}:
\begin{eqnarray}
\frac{ \left< V_{4} \, (V_{2}^{*})^{2} \, v_{2}^{~2} \right>^{A}} {  \left< V_{4}  \, (V_{2}^{*})^{2} \right>^{A}  \, \left<v_{2}^{~2} \right> } =   \frac{\langle \langle \cos (4\varphi_{1}^{A} + 2\varphi_{2}^{A} - 2\varphi_{3}^{B} - 2\varphi_{4}^{B} - 2\varphi_{5}^{B}) \rangle \rangle} {\langle \langle \cos (4\varphi_{1}^{A} - 2\varphi_{2}^{B} - 2\varphi_{3}^{B}) \rangle \rangle \, \langle  \langle \cos (2 \varphi_{1}^{A} - 2\varphi_{2}^{B}) \rangle \rangle}, \\
\frac{ \left< v_{2}^{~6} \right>} {  \left< v_{2}^{~4} \right>  \, \left<v_{2}^{~2} \right> } = \frac{\langle  \langle \cos (2\varphi_{1}^{A} + 2\varphi_{2}^{A} + 2\varphi_{3}^{A} - 2\varphi_{4}^{B} - 2\varphi_{5}^{B} - 2\varphi_{6}^{B} )\rangle \rangle }{\langle  \langle \cos (2\varphi_{1}^{A} + 2\varphi_{2}^{A} - 2\varphi_{3}^{B} - 2\varphi_{4}^{B})\rangle \rangle  \, \langle  \langle \cos (2 \varphi_{1}^{A} - 2\varphi_{2}^{B}) \rangle \rangle}.
\label{method:equality1}
\end{eqnarray}
Similarly, we can validate Eq.~(\ref{eq:equality2}) by calculating both sides with~\cite{Yan:2015jma}:
\begin{eqnarray}
\frac{ \left< V_{5} \, V_{3}^{*} \,  V_{2}^{*} \, v_{2}^{~2}  \right>^{A}} { \left< V_{5} \, V_{3}^{*} \,  V_{2}^{*}  \right>^{A} \, \left< v_{2}^{~2}\right>}  =  \frac{\langle \langle \cos (5\varphi_{1}^{A} + 2\varphi_{2}^{A} - 3\varphi_{3}^{B} - 2\varphi_{4}^{B} - 2\varphi_{5}^{B}) \rangle \rangle}{\langle \langle \cos (5\varphi_{1}^{A} - 3\varphi_{2}^{B} - 2\varphi_{3}^{B}) \rangle \rangle  \, \langle  \langle \cos (2 \varphi_{1}^{A} - 2\varphi_{2}^{B}) \rangle \rangle },\\
  \frac{ \left< v_{2}^{~4} \, v_{3}^{~2} \right>} {  \left< v_{2}^{~2} \, v_{3}^{~2} \right>  \, \left<v_{2}^{~2} \right> } = \frac{\langle  \langle \cos (3\varphi_{1}^{A} + 2\varphi_{2}^{A} + 2\varphi_{3}^{A} - 3\varphi_{4}^{B} - 2\varphi_{5}^{B} - 2\varphi_{6}^{B} )\rangle \rangle }{\langle  \langle \cos (3\varphi_{1}^{A} + 2\varphi_{2}^{A} - 3\varphi_{3}^{B} - 2\varphi_{4}^{B})\rangle \rangle  \, \langle  \langle \cos (2 \varphi_{1}^{A} - 2\varphi_{2}^{B}) \rangle \rangle}.
\label{method:equality2}
\end{eqnarray}
  
The magnitude of $V_{n}^{\rm NL}$ was denoted as $v_{n}\{\Psi_{m} \}$ (here $\Psi_{m}$ is the lower order flow symmetry plane and $m=2, 3$) in~\cite{Yan:2015jma}. The notation $v_{n,mk}$, where $n$ specifies the order of the flow term while $m$ and $k$ etc. denote the contributing lower order flow symmetry planes, is used in this Letter. If the linear and non-linear modes are independent, then the non-linear mode in higher order anisotropic flow can be analysed by correlating $V_{n}$ with $\Psi_{2}$ or/and $\Psi_3$~\cite{Yan:2015jma}. For sub-event A we can define:
\begin{eqnarray}
v_{4, 22}^{\,A} &=& \frac{\langle \langle \cos (4\varphi_{1}^{A} - 2\varphi_{2}^{B} - 2\varphi_{3}^{B}) \rangle \rangle} {\sqrt{\langle  \langle \cos (2\varphi_{1}^{A} + 2\varphi_{2}^{A} - 2\varphi_{3}^{B} - 2\varphi_{4}^{B})\rangle \rangle }} , \\ 
v_{5, 32}^{\,A} &=&  \frac{\langle \langle \cos (5\varphi_{1}^{A} - 3\varphi_{2}^{B} - 2\varphi_{3}^{B}) \rangle \rangle} {\sqrt{\langle  \langle \cos (3\varphi_{1}^{A} + 2\varphi_{2}^{A} - 3\varphi_{3}^{B} - 2\varphi_{4}^{B})\rangle \rangle }} , \\ 
v_{6, 222}^{\,A} &=& \frac{\langle \langle \cos (6\varphi_{1}^{A} - 2\varphi_{2}^{B} - 2\varphi_{3}^{B} - 2\varphi_{4}^{B}) \rangle \rangle} {\sqrt{\langle  \langle \cos (2\varphi_{1}^{A} + 2\varphi_{2}^{A} + 2\varphi_{3}^{A} - 2\varphi_{4}^{B} - 2\varphi_{5}^{B} - 2\varphi_{6}^{B} )\rangle \rangle }} ,\\ 
v_{6, 33}^{\,A} &=& \frac{\langle \langle \cos (6\varphi_{1}^{A} - 3\varphi_{2}^{B} - 3\varphi_{3}^{B}) \rangle \rangle} {\sqrt{\langle  \langle \cos (3\varphi_{1}^{A} + 3\varphi_{2}^{A} - 3\varphi_{3}^{B} - 3\varphi_{4}^{B})\rangle \rangle }}.
\label{eq:V2nApsin}
\end{eqnarray}
Similarly, one can obtain $v_{n,mk}^{B}$ for sub-event B. The average of $v_{n,mk}^{A}$ and $v_{n,mk}^{B}$, defined as $v_{n,mk}$, quantifies the magnitude of the non-linear mode in higher order anisotropic flow, which can be written as~\cite{Bhalerao:2013ina}:
\begin{eqnarray}
v_{4, 22} &=& \frac{\langle v_{4} \, v_{2}^{2} \, \cos (4\Psi_{4} - 4\Psi_{2}) \rangle} {\sqrt{\langle  v_{2}^{4} \rangle }} \approx \langle v_{4} \, \cos (4\Psi_{4} - 4\Psi_{2}) \rangle, \\ 
v_{5, 32} &=& \frac{\langle v_{5} \, v_{3} \, v_{2} \, \cos (5\Psi_{5} - 3\Psi_{3} - 2\Psi_{2}) \rangle} {\sqrt{\langle  v_{3}^{2} \, v_{2}^{2}  \rangle }} \approx \langle v_{5} \, \cos (5\Psi_{5} - 3\Psi_{3} - 2\Psi_{2})  \rangle,  \label{eq:v532}\\
v_{6, 222} &=& \frac{\langle v_{6} \, v_{2}^{3} \, \cos (6\Psi_{6} - 6\Psi_{2}) \rangle} {\sqrt{\langle  v_{2}^{6} \rangle }} \approx \langle v_{6} \, \cos (6\Psi_{6} - 6\Psi_{2}) \rangle , \\ 
v_{6, 33} &=& \frac{\langle v_{6} \, v_{3}^{2} \, \cos (6\Psi_{6} - 6\Psi_{3}) \rangle} {\sqrt{\langle  v_{3}^{4} \rangle }} \approx \langle v_{6} \, \cos (6\Psi_{6} - 6\Psi_{3}) \rangle.  
\label{eq:V2nApsin2}
\end{eqnarray}
The approximation is valid if the correlation between lower ($n=2,3$) and higher ($n>3$) flow coefficients is weak.

As can be seen in Eqs.~(\ref{eq:V4}-\ref{eq:V6}), the calculation for $V_{6}$ is more complicated than $V_{4}$ and $V_{5}$, and the exact expression for $v_{6}^{\rm L}$ is currently not available. Therefore, we only focus on the two non-linear modes of $V_{6}$ without discussing $v_{6}^{\rm L}$. According to Eqs.~(\ref{eq:V4}) to (\ref{eq:V5}), the magnitudes of the linear mode in higher order anisotropic flow can be calculated as:
\begin{eqnarray}
v_{4}^{\rm L} = \sqrt{v_{4}^{\,2}\{2\} - v_{4, 22}^{\,2}}, \\
v_{5}^{\rm L} = \sqrt{v_{5}^{\,2}\{2\} - v_{5, 32}^{\,2}}.
\label{eq:vnPsim}
\end{eqnarray}

The ratio of $v_{n,mk}$ to $v_{n}\{2\}$, denoted as $\rho_{n,mk}$, can be calculated as:
\begin{eqnarray}
\rho_{4,22} &=& \frac{v_{4, 22}}{v_{4}\{2\}}   = \langle  \cos (4\Psi_{4} - 4 \Psi_{2}) \rangle, \\ 
\rho_{5,32} &=& \frac{v_{5, 32}}{v_{5}\{2\}}  = \langle  \cos (5\Psi_{5} - 3\Psi_{3} - 2\Psi_{2}) \rangle,  \label{eq:rho532}\\ 
\rho_{6,222} &=& \frac{v_{6, 222}}{v_{6}\{2\}} = \langle \cos (6\Psi_{6} - 6\Psi_{2}) \rangle, \\ 
\rho_{6,33} &=& \frac{v_{6, 33}}{v_{6}\{2\}} = \langle \cos (6\Psi_{6} - 6\Psi_{3}) \rangle. 
\label{eq:V2npsin}
\end{eqnarray}
These observables measure the correlations between different order flow symmetry planes if the correlations between different order flow coefficients are weak. They are very similar to the so-called weighted event-plane correlations measured by the ATLAS Collaboration~\cite{Aad:2014fla}. The differences are as follows: $\langle v_{2}^{2} \, v_{3}^{2} \rangle$ is used in Eq.~(\ref{eq:v532}) and (\ref{eq:rho532}), while $\langle v_{2}^{2} \rangle \langle v_{3}^{2} \rangle$ was used in~\cite{Aad:2014fla}, which did not consider the anti-correlations between $v_{2}$ and $v_{3}$ found in~\cite{ALICE:2016kpq}. In addition, multi-particle correlations are used for $v_2$ and $v_3$ in the denominator of the observables, while two-particle correlations are used in the event-plane correlations which might be biased from fluctuations of $v_{2}$ and $v_3$.

The non-linear mode coefficients $\chi_{n,mk}$ in Eqs.~(\ref{eq:V4}) to (\ref{eq:V6}) are defined as:
\begin{eqnarray}
\chi_{4,22} &=& \frac{v_{4, 22}} {\sqrt{\langle v_{2}^{4} \rangle}}\\ 
\chi_{5,32} &=&  \frac{v_{5, 32}} {\sqrt{\langle v_{2}^{2} \, v_{3}^{2} \rangle}}\label{eq:chi_532}\\ 
\chi_{6,222} &=&  \frac{v_{6, 222}} {\sqrt{\langle v_{2}^{6} \rangle}}\\ 
\chi_{6,33} &=& \frac{v_{6, 33}} {\sqrt{\langle v_{3}^{4} \rangle}}.
\end{eqnarray}
\label{eq:chi_n}
These quantify the contributions of the non-linear mode and are expected to be independent of $v_{2}$ or $v_{3}$.

All of the observables above are based on 2- and multi-particle correlations, which can be obtained using the generic framework for anisotropic flow analyses introduced in Ref.~\cite{Bilandzic:2013kga}. 


\section{Experimental setup and data analysis}
\label{sec:setup}

The data samples analysed in this Letter were recorded by ALICE during the Pb--Pb runs of the LHC at a centre-of-mass energy of $\snn=2.76$ TeV in 2010. Minimum bias Pb--Pb collision events were triggered by the coincidence of signals in the V0 detector~\cite{Aamodt:2008zz, Abbas:2013taa}, with an efficiency of 98.4\% of the hadronic cross section~\cite{Abelev:2013qoq}. The V0 detector is composed of two arrays of scintillator counters, V0-A and V0-C, which cover the pseudorapidity ranges $2.8 < \eta < 5.1$ and $-3.7 < \eta < -1.7$, respectively. Beam background events were rejected using the timing information from the V0 and the Zero Degree Calorimeter (ZDC)~\cite{Aamodt:2008zz} detectors and by correlating reconstructed clusters and tracklets with the Silicon Pixel Detectors (SPD). The fraction of pile-up events in the data sample is found to be negligible after applying dedicated pile-up removal criteria~\cite{Abelev:2014ffa}. Only events with a reconstructed primary vertex within $\pm 10$ cm from the nominal interaction point along the beam direction were used in this analysis. The primary vertex was estimated using tracks reconstructed by the Inner Tracking System (ITS)~\cite{Aamodt:2008zz, Aamodt:2010aa} and Time Projection Chamber (TPC)~\cite{Aamodt:2008zz, Alme:2010ke}. The collision centrality was determined from the measured V0 amplitude and centrality intervals were defined following the procedure described in~\cite{Abelev:2013qoq}. About 13 million Pb--Pb events passed all of the event selection criteria. 

Tracks reconstructed using the combined information from the TPC and ITS are used in this analysis. This combination ensures a high detection efficiency, optimum momentum resolution, and a minimum contribution from photon conversions and secondary charged particles produced either in the detector material or from weak decays. To reduce the contributions from secondaries, charged tracks were required to have a distance of closest approach to the primary vertex in the longitudinal ($z$) direction and transverse ($xy$) plane smaller than 3.2 cm and 2.4 cm, respectively. Additionally, tracks were required to have at least 70 TPC space points out of the maximum 159. The average $\chi^{2}$ per degree of freedom of the track fit to the TPC space points was required to be below 2. In this study, tracks were selected in the pseudorapidity range $|\eta|<0.8$ and the transverse momentum range 0.2 $< \pt <$ 5.0~GeV/$c$.


\section{Systematic uncertainties}
\label{sec:sys}


Numerous sources of systematic uncertainty were investigated by varying the event and track selection as well as the uncertainty associated with the possible remaining non-flow effects in the analysis. The variation of the results with the collision centrality is calculated by alternatively using the TPC or SPD to estimate the event multiplicity and is found to be less than 3\% for all observables. Results with opposite polarities of the magnetic field within the ALICE detector and with narrowing the nominal $\pm$10 cm range of the reconstructed vertex along the beam direction from the centre of the ALICE detector to 9, 8 and 7 cm do not show a difference of more than 2\% compared to the default selection criteria for various measurements. The contributions from pileup events to the final systematic uncertainty are found to be negligible. 
The sensitivity to the track selection criteria was explored by varying the number of TPC space points and by using tracks reconstructed in the TPC alone.  Varying the number of TPC space points from 70 to 80, 90 and 100 out of a possible 158, results in a 1--3\% variation of the results for $v_{n}$, within 1.5\% for $\rho_{n,mk}$ and $\chi_{n,mk}$. Using TPC-only tracks leads to a difference of less than 14\%, 17\% and 8\% for $v_{n}$, $\rho_{n,mk}$ and $\chi_{n,mk}$, respectively. Both effects were included in the evaluation of the systematic uncertainty.
Several different approaches have been applied to estimate the effects of non-flow. These include the investigation of multi-particle correlations with various $|\Delta \eta|$ gaps, the application of the like-sign technique which correlates two particles with either all positive or negative charges and suppress such non-flow as due to resonance decays, as well as the calculations using HIJING Monte Carlo simulations~\cite{Gyulassy:1994ew}, which do not include anisotropic flow. It was found that the possible remaining non-flow effects are less than 10.5\%, 11\% and 7\% for $v_{n}$, $\rho_{n,mk}$ and $\chi_{n,mk}$, respectively. They are taken into account in the final systematic uncertainty. The systematic uncertainties evaluated for each source mentioned above were added in quadrature to obtain the total systematic uncertainty of the measurements.

\section{Results and discussion}
{\label{sec:Results}


\begin{figure}[tbh]
\begin{center}
\includegraphics[width=0.88\textwidth]{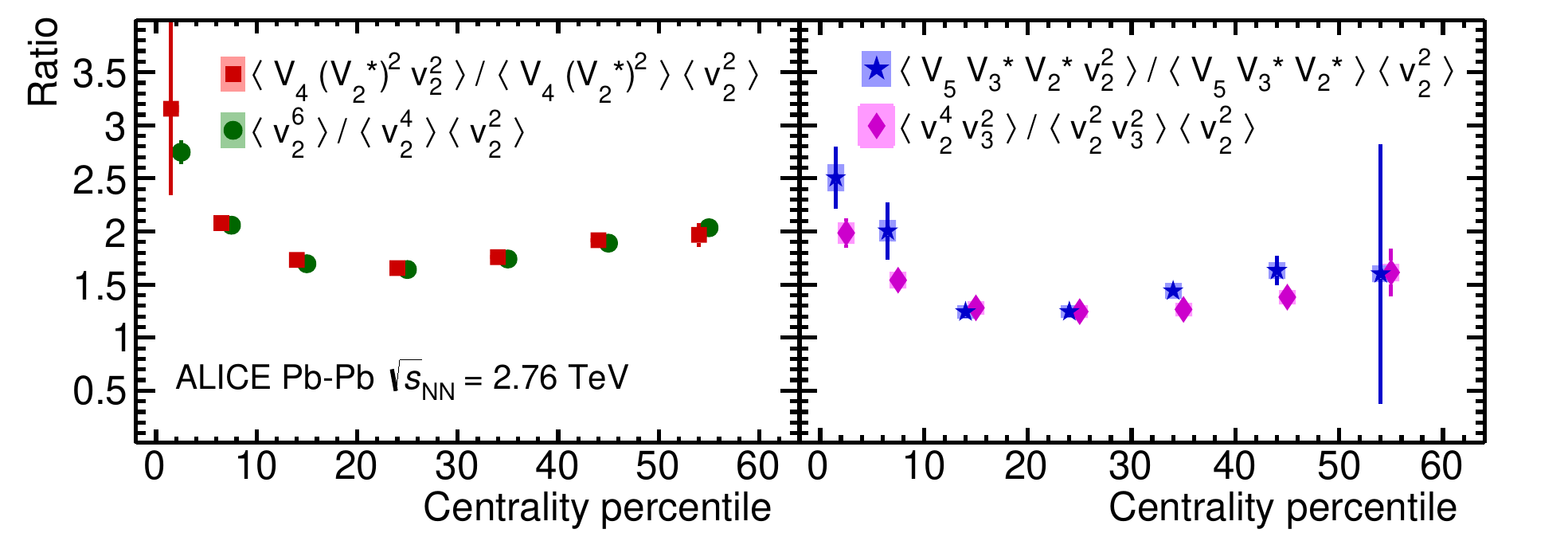}
\caption{Study of relationship between linear and non-linear modes in higher order anisotropic flow in Pb--Pb collisions at $\sqrt{s_{_{\rm NN}}} = 2.76$ TeV, according to Eqs.~(\ref{eq:equality1}) and (\ref{eq:equality2}). }
\label{fig:test} 
\end{center}
\end{figure}

As discussed in Sec.~\ref{sec:probe}, one can validate the assumption that linear and non-linear modes in higher order anisotropic flow are uncorrelated via Eqs.~(\ref{eq:equality1}) and (\ref{eq:equality2}).
These have been tested in A Multi-Phase Transport (AMPT) model~\cite{Bhalerao:2014xra} as well as in the hydrodynamic calculations~\cite{Qian:2016fpi}. Good agreement between left- and right-hand sides of Eqs.~(\ref{eq:equality1}) and (\ref{eq:equality2}) is found for all centrality classes, independent of the initial conditions and the ideal or viscous fluid dynamics used in the calculations. Thus, it is crucial to check these equalities in data, to further confirm the assumption that the two components are uncorrelated and can be isolated independently. Figure~\ref{fig:test} confirms that the agreement observed in theoretical calculations is also present in the data despite small deviations found in central collisions when testing Eqs.~(\ref{eq:equality2}). Their centrality dependency are similar as the previous theoretical predictions~\cite{Bhalerao:2014xra, Qian:2016fpi}. The measurements support the assumption that higher order anisotropic flow $V_{n}$ ($n>3$) can be modeled as the sum of independent linear and non-linear modes.


The magnitudes of linear and non-linear modes in higher order anisotropic flow are reported as a function of collision centrality in Fig.~\ref{fig:vnl}. In this Letter, sub-events A and B are built in the pseudorapidity ranges $-0.8 < \eta < -0.4$ and $0.4<\eta<0.8$, respectively, which results in a pseudorapidity gap of $|\Delta\eta|>0.8$ for all presented measurements. It can be seen that the linear mode $v_{4}^{\rm L}$ depends weakly on centrality and is the larger contribution to $v_{4}\{2\}$ for the centrality range 0--30\%. The non-linear mode, $v_{4, 22}$, increases monotonically as the centrality decreases and saturates around centrality percentile 50\%, becoming the dominant source for centrality intervals above 40\%. Similar trends of centrality dependence have been observed for $V_{5}$, although $v_{5,32}$ becomes the dominant contribution in centrality percentile above 30\%. Only two non-linear components $v_{6, 222}$ and $v_{6, 33}$ are discussed for $V_{6}$. It is shown in Fig.~\ref{fig:vnl} (right) that $v_{6, 222}$ increases monotonically as the centrality decreases to centrality 50\%, while $v_{6, 33}$ has a weaker centrality dependence compared to $v_{6, 222}$. 

The linear and non-linear modes in higher order anisotropic flow were investigated by the ATLAS Collaboration~\cite{Aad:2015lwa} using a different approach based on ``Event Shape Engineering''~\cite{Schukraft:2012ah}. With this method one can utilise large fluctuations in the initial geometry of the system to select events corresponding to a specific initial shape. The conclusion is qualitatively consistent with what is reported here, although a direct comparison is not possible due to the different kinematic cuts (especially the integrated $\pt$ range) used in the two measurements. The higher order anisotropic flow induced by lower order anisotropic flow were also measured using the event-plane method at the LHC~\cite{Abelev:2012di,Chatrchyan:2013kba}. However, the measurements of the non-linear mode presented in this Letter are based on the multi-particle correlations method with a $|\Delta \eta|$ gap. This method makes it easier to measure an observable like $v_{5, 32}$, which is less straightforward to define using the event plane method~\cite{Abelev:2012di, Chatrchyan:2013kba}. In addition, as pointed out in~\cite{Luzum:2012da,Yan:2015jma,Bhalerao:2014xra}, this new multi-particle correlations method should strongly suppress short-range (in pseudorapidity) non-flow effects and provides a robust measurement without any dependence on the experimental acceptance. The measurements are compared to recent hydrodynamic calculations from a hybrid ${\tt IP}$-${\tt Glasma+MUSIC+UrQMD}$ model~\cite{McDonald:2016vlt}, in which realistic event-by-event initial conditions are used and the hydrodynamic evolution takes into account both shear and bulk viscosity. It is shown that this hydrodynamic calculation could describe quantitatively the total magnitudes of $V_{4}$ and $V_{6}$, as well as the magnitudes of their linear and non-linear modes, while it slightly overestimates the results for $V_{5}$. 

\begin{figure}[tbh]
\begin{center}
\includegraphics[width=0.95\textwidth]{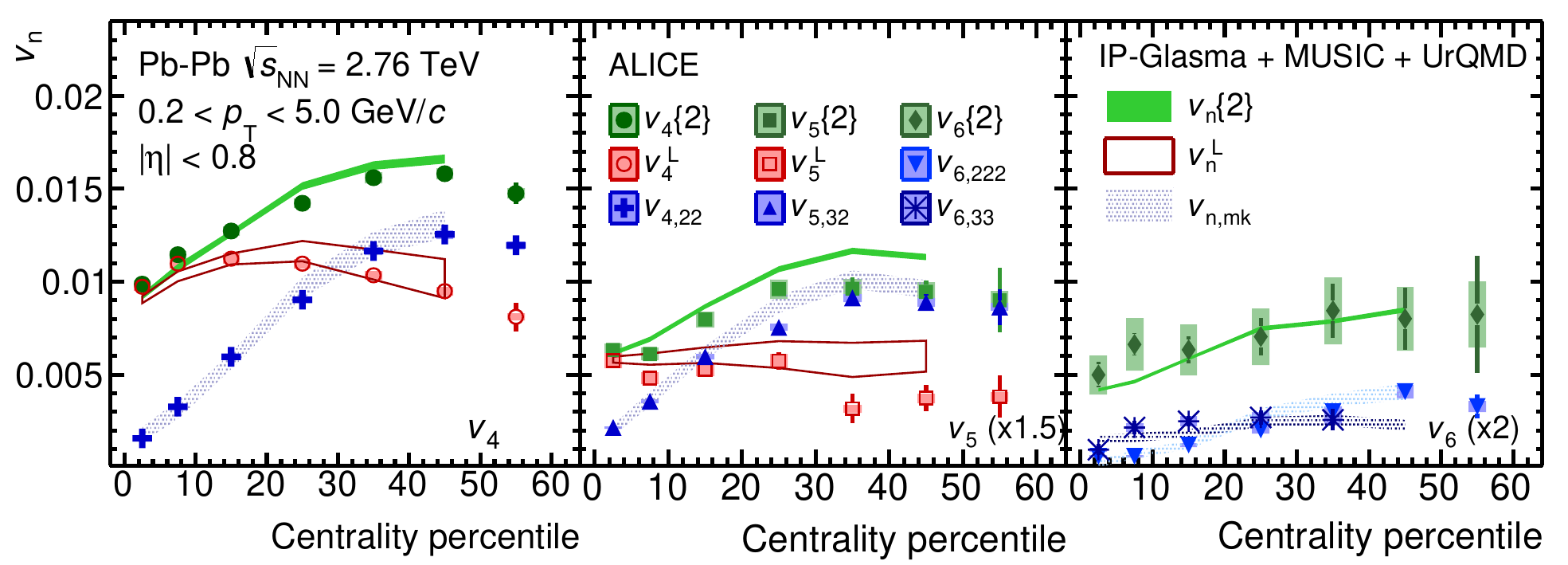}
\caption{Centrality dependence of $v_4$ (left), $v_5$ (middle) and $v_6$ (right) in Pb--Pb collisions at $\sqrt{s_{_{\rm NN}}} = 2.76$ TeV. Contributions from linear and non-linear modes are presented with open and solid markers, respectively. The hydrodynamic calculations from ${\tt IP}$-${\tt Glasma+MUSIC+UrQMD}$~\cite{McDonald:2016vlt} are shown for comparison.}
\label{fig:vnl} 
\end{center}
\end{figure}


\begin{figure}[tbh]
\begin{center}
\includegraphics[width=0.98\textwidth]{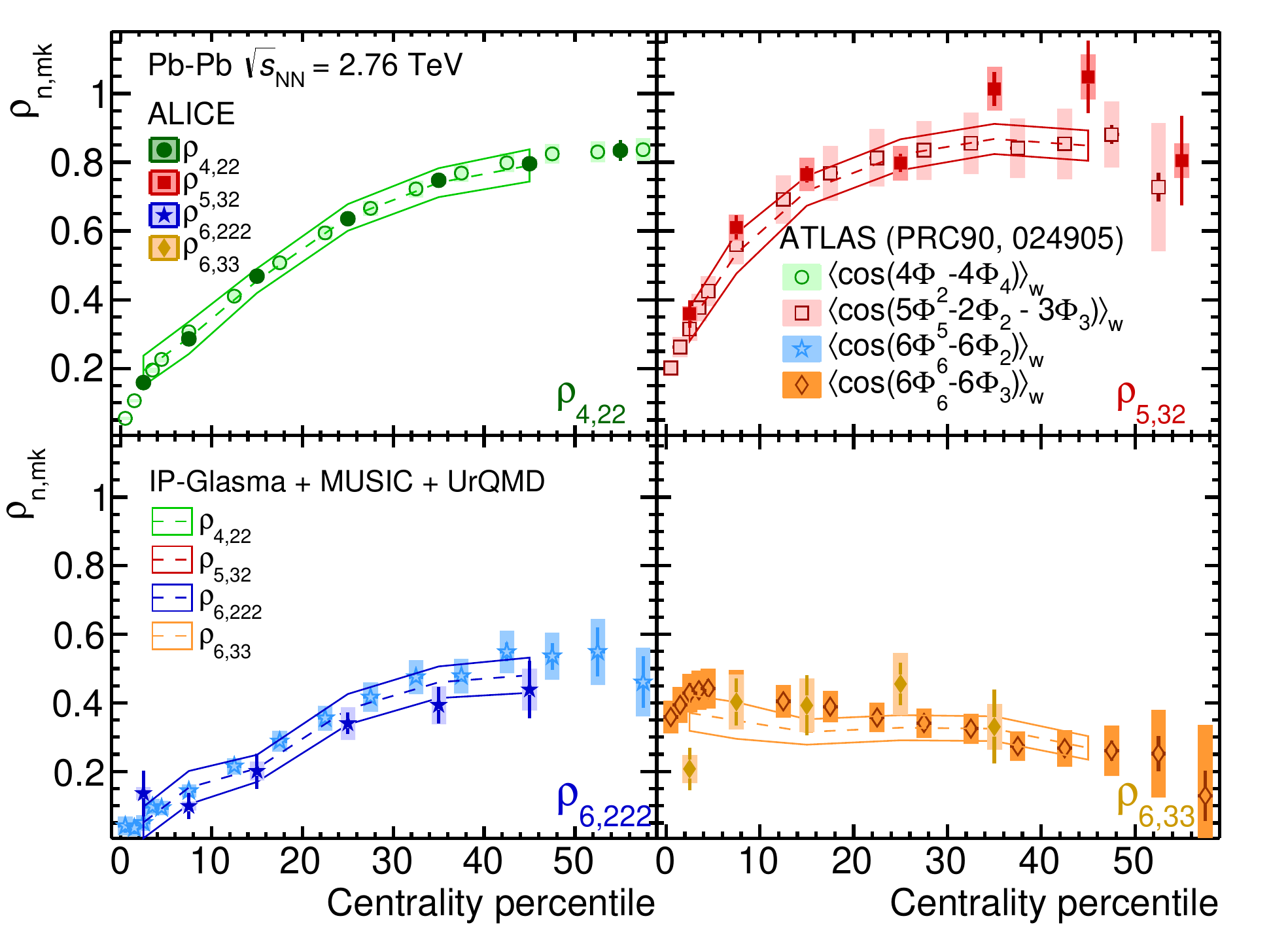}
\caption{Centrality dependence of $\rho_{n,mk}$ in Pb--Pb collisions at $\sqrt{s_{_{\rm NN}}} = 2.76$ TeV. ATLAS measurements based on the event-plane correlation~\cite{Aad:2014fla} are presented with open markers. The hydrodynamic calculations from ${\tt IP}$-${\tt Glasma+MUSIC+UrQMD}$~\cite{McDonald:2016vlt} are shown with open bands. }
\label{fig:rho} 
\end{center}
\end{figure}

The centrality dependence of $\rho_{n,mk}$, which quantifies the angular correlations between different order flow symmetry planes, is presented in Fig.~\ref{fig:rho}. It is observed that $\rho_{4,22}$ increases from central to peripheral collisions, which suggests that the correlations between $\Psi_{2}$ and $\Psi_{4}$ are stronger in peripheral than in central collisions. It implies that $V_{4}^{\rm NL}$ tends to align with $V_{4}$ in more peripheral collisions. The results of $\rho_{6,33}$, which measures the correlation of $\Psi_{3}$ and $\Psi_{6}$, do not exhibit a strong centrality dependence within the statistical uncertainties. As mentioned above, $\rho_{4,22}$ and $\rho_{6,33}$ are similar to the previous ``event-plane correlation'' measurements $\left< \cos (4\Phi_{4} - 4\Phi_{2}) \right>_{w}$ and $\left< \cos (6\Phi_{6} - 6\Phi_3) \right>_{w}$ in~\cite{Aad:2014fla}. The comparisons between measurements of these observables are also presented in Fig.~\ref{fig:rho}. The results are compatible with each other, despite the different kinematic ranges used by ATLAS and this analysis. It should be also noted that the measurements of $\rho_{n,mk}$ presented in this Letter show the symmetry plane correlations at mid-pseudorapidity $|\eta|<0.8$ while ATLAS measured the symmetry plane correlations using $-4.8 < \eta < -0.5$ and $0.5 < \eta < 4.8$ for two-plane correlations, and using $-2.7 < \eta < -0.5$, $0.5 < \eta < 2.7$ and $3.3 < |\eta| < 4.8$ for 3-plane correlations~\cite{Aad:2014fla}. Previous investigations suggest that there might be $\eta$-dependent fluctuations of the flow symmetry plane and the flow magnitude~\cite{Khachatryan:2015oea,Pang:2014pxa}. As a consequence, one might expect a difference when measuring the correlations of flow symmetry planes from different pseudorapidity regions. However, Fig.~\ref{fig:rho} shows good agreement between the ALICE and ATLAS measurements. Therefore, no obvious indication that the flow symmetry plane varies with $\eta$ can be deduced from this comparison. It is noticeable in Fig.~\ref{fig:rho} that the $\rho_{5,32}$ measurement seems slightly higher than the $\left< \cos (5\Phi_{5} - 3\Phi_{3} - 2\Phi_{2}) \right>_{w}$ measurement. This is mainly due to a small difference between the definitions of the observable as introduced in Sec.~\ref{sec:probe}: the term $\left< v_{2}^{2} \, v_{3}^{2} \right>^{1/2} $ is used in $\rho_{5,32}$, whereas $\left< v_{2}^{2} \right>^{1/2} \left< v_{3}^{2} \right>^{1/2}$ is used in the ``event-plane correlations''~\cite{Aad:2014fla}. Considering the known anti-correlations between $v_{2}$ and $v_{3}$~\cite{Aad:2015lwa, ALICE:2016kpq}, $\left< v_{2}^{2} \, v_{3}^{2} \right>^{1/2}$ could be up to 10\% lower than $\left< v_{2}^{2} \right>^{1/2}\left< v_{3}^{2} \right>^{1/2}$ depending on the centrality class~\cite{ALICE:2016kpq}, leading to a slightly larger $\rho_{5,32}$ than $\left< \cos (5\Phi_{5} - 3\Phi_{3} - 2\Phi_{2}) \right>_{w}$ from ATLAS. 

It has been observed in hydrodynamic and transport model calculations that the symmetry plane correlations, e.g. correlations of second and fourth order symmetry planes, change sign during the system evolution~\cite{Qiu:2012uy, Teaney:2013dta, Zhou:2015eya}. The measured flow symmetry plane correlations could be nicely explained by the combination of contributions from linear and non-linear modes in higher order anisotropic flow~\cite{Teaney:2013dta}. This indicates that the flow symmetry plane correlation $\rho_{n,mk}$ carries important information about the dynamic evolution of the created system. In addition, the model calculations suggest that stronger initial symmetry plane correlations are reflected in stronger correlations between the flow symmetry planes in the final state~\cite{Qiu:2012uy,Zhou:2015eya}. And a larger value of $\eta/s$ of the QGP leads to weaker flow symmetry plane correlations in the final state. As pointed out in~\cite{Qiu:2012uy}, the hydrodynamic calculations from ${\tt VISH2+1}$ using Monte Carlo Glauber (MC-Glb) or Monte Carlo Kharzeev-Levin-Nardi (MC-KLN) initial conditions can only describe qualitatively the trends of the centrality dependence of the event-plane correlation measurements by ATLAS. It is therefore expected that these hydrodynamic calculations cannot describe the presented ALICE measurements, which are compatible with the ATLAS event-plane correlation measurements. Figure~\ref{fig:rho} shows that the hydrodynamic calculations from ${\tt IP}$-${\tt Glasma+MUSIC+UrQMD}$~\cite{McDonald:2016vlt} reproduce nicely the measurements of symmetry plane correlations $\rho_{n,mk}$. The measurements of $\rho_{n,mk}$ presented in this Letter, together with the comparison to hydrodynamic calculations, should place constraints on the initial conditions and $\eta/s$ of the QGP in hydrodynamic calculations.



\begin{figure}[tbh]
\begin{center}
\includegraphics[width=0.98\textwidth]{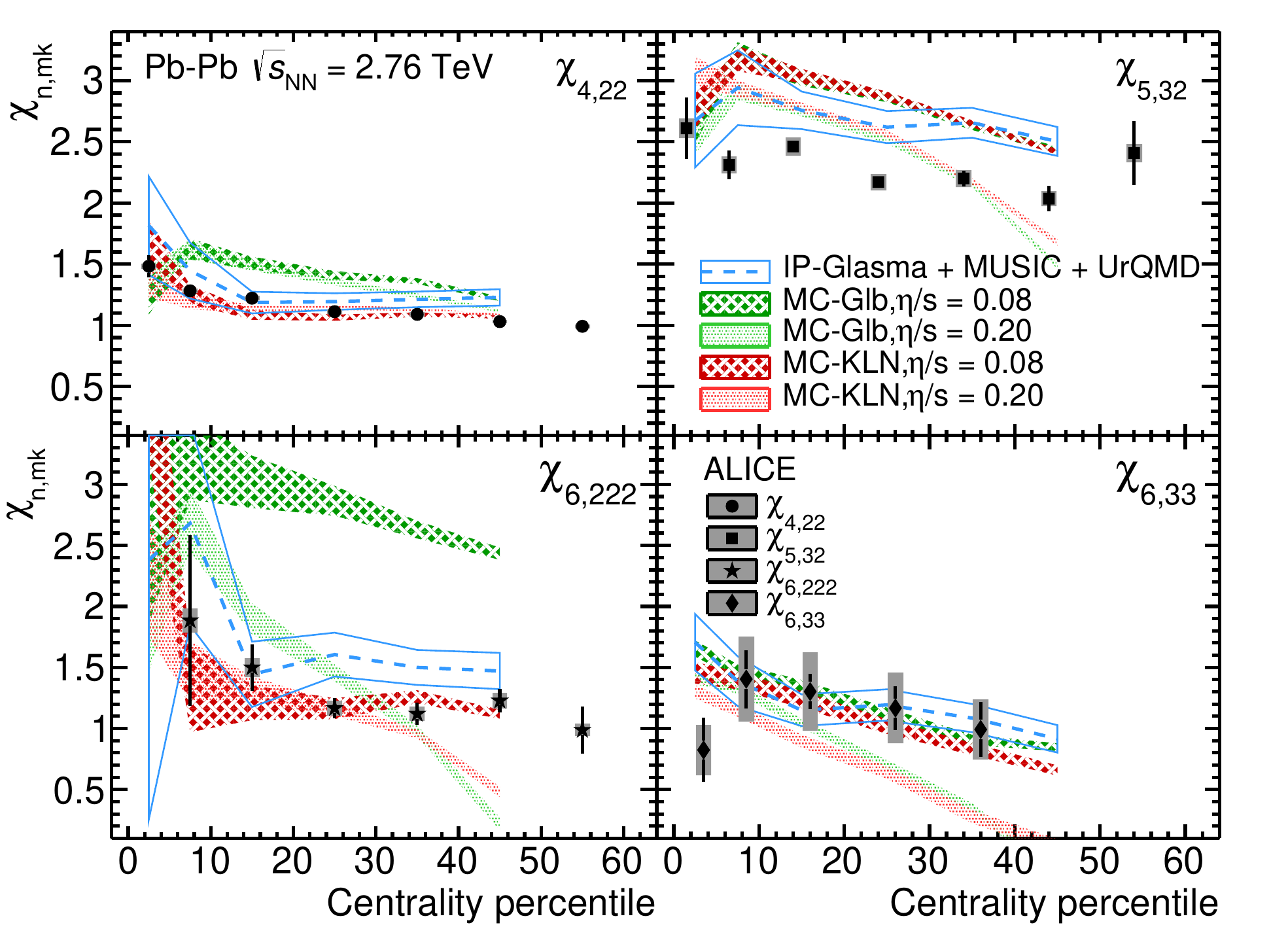}
\caption{Centrality dependence of $\chi$ in Pb--Pb collisions at $\sqrt{s_{_{\rm NN}}} = 2.76$ TeV. Hydrodynamic calculations from ${\tt VISH2+1}$~\cite{Qian:2016fpi} are shown in shaded areas and the one from ${\tt IP}$-${\tt Glasma+MUSIC+UrQMD}$~\cite{McDonald:2016vlt} are shown with open bands. }
\label{fig:chi} 
\end{center}
\end{figure}

Figure~\ref{fig:chi} presents the measurements of the non-linear mode coefficients as a function of collision centrality. It is observed that $\chi_{4,22}$ and $\chi_{6,222}$ decrease modestly from central to mid-central collisions, and stay almost constant from mid-central to more peripheral collisions. For $\chi_{5,32}$ and $\chi_{6,33}$ strong centrality dependence is not observed either. Thus, the dramatic increase of $v_{n,mk}$ shown in Fig.~\ref{fig:vnl} appears to be mainly due to the increase of $v_{2}$ and/or $v_{3}$ from central to peripheral collisions and not the increase of the non-linear mode coefficient. It is also noteworthy that the relationship of $\chi_{4,22} \sim \chi_{6,33} \approx \frac{\chi_{5,32}}{2}$ is approximately valid, as predicted by hydrodynamic calculations~\cite{Yan:2015jma}. The comparisons to event-by-event viscous hydrodynamic calculations from ${\tt VISH2+1}$~\cite{Qian:2016fpi}  and from ${\tt IP}$-${\tt Glasma+MUSIC+UrQMD}$~\cite{McDonald:2016vlt} are also presented in Fig.~\ref{fig:chi}. ${\tt VISH2+1}$ shows that $\chi_{4,22}$ calculations with MC-Glb initial conditions are larger than those with MC-KLN initial conditions, i.e. $\chi_{4,22}$ depends on the initial conditions. At the same time, the curves with different $\eta/s$ values for ${\tt VISH2+1}$ are very similar, indicating that $\chi_{4,22}$ is insensitive to $\eta/s$. The measurements favor IP-Glasma and MC-KLN over MC-Glb initial conditions regardless of $\eta/s$. This suggests that the $\chi_{4,22}$ measurement can be used to constrain the initial conditions, with less concern of the setting of $\eta/s(T)$ in hydrodynamic calculations than previous flow observables. 
 
It was predicted that $\chi_{6,222} < \chi_{6,33}$ based on the ideal hydrodynamic calculation using smooth initial Gaussian density profiles~\cite{Yan:2015jma}, whereas an opposite prediction was obtained in the ideal hydrodynamic calculation evolving genuinely bumpy initial conditions obtained from a Monte Carlo sampling of the initial nucleon positions in the colliding nuclei~\cite{Qian:2016fpi}. It is seen in Fig.~\ref{fig:chi} that $\chi_{6,222} \sim \chi_{6,33}$ within the current uncertainties. The data are not able to discriminate the different predictions in~\cite{Yan:2015jma} and~\cite{Qian:2016fpi}. Hydrodynamic calculations using MC-KLN and IP-Glasma initial conditions give better descriptions of $\chi_{6,222}$, compared to the ones using MC-Glb initial conditions. For $\chi_{5,32}$ none of the combinations of initial conditions and $\eta/s$ in the hydrodynamic calculations agree quantitatively with data. This might be due to the current difficulty of describing the anti-correlations between $v_{2}$ and $v_{3}$ in hydrodynamic calculations~\cite{ALICE:2016kpq,Zhu:2016puf}, which are involved in the calculation of $\chi_{5,32}$. Furthermore, ${\tt VISH2+1}$ calculations show that $\chi_{5,32}$ and $\chi_{6,33}$ are very weakly sensitive to the initial conditions, but decrease as $\eta/s$ increases. The investigation with the ${\tt VISH2+1}$ hydrodynamic framework shows that the sensitivity of $\chi_{5,32}$ and $\chi_{6,33}$ to $\eta/s$ is not due to sensitivity to shear viscous effects during the buildup of hydrodynamic flow. Instead, as found in~\cite{Qian:2016fpi}, it is due to the $\eta/s$ at freeze-out. The measurements of $\chi_{5,32}$ and $\chi_{6,33}$ do not further constrain the $\eta/s$ during system evolution, however, they provide unique information on $\eta/s$ at freeze-out which was poorly known and cannot be obtained from other anisotropic flow related observables. Further improvement of model calculations on correlations between different order flow coefficients are necessary to better understand the comparison of $\chi_{5,32}$ obtained from data and hydrodynamic calculations. The $\chi_{6,33}$ results are consistent with hydrodynamic calculations from ${\tt VISH2+1}$ with MC-KLN initial conditions using $\eta/s=0.08$ and ${\tt IP}$-${\tt Glasma+MUSIC+UrQMD}$ with a $\eta/s=0.095$. It is shown that $\chi_{5,32}$ and $\chi_{6,33}$ have a weak centrality dependence if a smaller $\eta/s$ is used in the hydrodynamic calculations. Such a weak centrality dependence of $\chi_{5,32}$ and $\chi_{6,33}$ is observed in data as well. The measurements presented here suggest a small $\eta/s$ value at freeze-out, which can be useful to constrain the temperature dependence of the shear viscosity over entropy density ratio, $\eta/s(\rm T)$, in the development of hydrodynamic frameworks. These results suggest that future tuning of the parameterisations of $\eta/s(\rm T)$ in hydrodynamic frameworks using the presented measurements is necessary.

\section{Summary}
\label{sec:summary}

The linear and non-linear modes in higher order anisotropic flow generation were studied with 2- and multi-particle correlations in Pb--Pb collisions at $\sqrt{s_{_{\rm NN}}}=$ 2.76 TeV. The results presented in this Letter show that higher order anisotropic flow can be isolated into two independent contributions: the component that arises from a non-linear response of the system to the lower order initial anisotropy coefficients $\varepsilon_{2}$ and/or $\varepsilon_{3}$, and a linear mode which is driven by linear response of the system to the same order cumulant-defined anisotropy coefficient. A weak centrality dependence is observed for the contributions from linear mode whereas the contributions from non-linear mode increase dramatically as the collision centrality decreases, and it becomes the dominant source in higher order anisotropic flow in mid-central to peripheral collisions. It is shown that this is mainly due to the increase of lower order flow coefficients $v_{2}$ and $v_{3}$. The correlations between different flow symmetry planes are measured. The results are compatible with the previous ``event-plane correlation'' measurements, and can be quantitatively described by  calculations using the ${\tt IP}$-${\tt Glasma+MUSIC+UrQMD}$ framework. Furthermore, non-linear mode coefficients, which have different sensitivities to the shear viscosity over entropy density ratio $\eta/s$ and the initial conditions, are presented in this Letter. Comparisons to hydrodynamic calculations suggest that the data is described better by hydrodynamic calculations with smaller $\eta/s$. In addition, the MC-Glb initial condition is disfavoured by the presented results.

Measurements of linear and non-linear modes in higher order anisotropic flow and their comparison to hydrodynamic calculations provide more precise constraints on the initial conditions and temperature dependence of $\eta/s$.
These results could also offer new insights into the geometry of the fluctuating initial state and into the dynamical evolution of the strongly interacting medium produced in relativistic heavy-ion collisions at the LHC.

\newenvironment{acknowledgement}{\relax}{\relax}
\begin{acknowledgement}
\section*{Acknowledgements}

The ALICE Collaboration would like to thank all its engineers and technicians for their invaluable contributions to the construction of the experiment and the CERN accelerator teams for the outstanding performance of the LHC complex.
The ALICE Collaboration gratefully acknowledges the resources and support provided by all Grid centres and the Worldwide LHC Computing Grid (WLCG) collaboration.
The ALICE Collaboration acknowledges the following funding agencies for their support in building and running the ALICE detector:
A. I. Alikhanyan National Science Laboratory (Yerevan Physics Institute) Foundation (ANSL), State Committee of Science and World Federation of Scientists (WFS), Armenia;
Austrian Academy of Sciences and Nationalstiftung f\"{u}r Forschung, Technologie und Entwicklung, Austria;
Ministry of Communications and High Technologies, National Nuclear Research Center, Azerbaijan;
Conselho Nacional de Desenvolvimento Cient\'{\i}fico e Tecnol\'{o}gico (CNPq), Universidade Federal do Rio Grande do Sul (UFRGS), Financiadora de Estudos e Projetos (Finep) and Funda\c{c}\~{a}o de Amparo \`{a} Pesquisa do Estado de S\~{a}o Paulo (FAPESP), Brazil;
Ministry of Science \& Technology of China (MSTC), National Natural Science Foundation of China (NSFC) and Ministry of Education of China (MOEC) , China;
Ministry of Science, Education and Sport and Croatian Science Foundation, Croatia;
Ministry of Education, Youth and Sports of the Czech Republic, Czech Republic;
The Danish Council for Independent Research | Natural Sciences, the Carlsberg Foundation and Danish National Research Foundation (DNRF), Denmark;
Helsinki Institute of Physics (HIP), Finland;
Commissariat \`{a} l'Energie Atomique (CEA) and Institut National de Physique Nucl\'{e}aire et de Physique des Particules (IN2P3) and Centre National de la Recherche Scientifique (CNRS), France;
Bundesministerium f\"{u}r Bildung, Wissenschaft, Forschung und Technologie (BMBF) and GSI Helmholtzzentrum f\"{u}r Schwerionenforschung GmbH, Germany;
General Secretariat for Research and Technology, Ministry of Education, Research and Religions, Greece;
National Research, Development and Innovation Office, Hungary;
Department of Atomic Energy Government of India (DAE) and Council of Scientific and Industrial Research (CSIR), New Delhi, India;
Indonesian Institute of Science, Indonesia;
Centro Fermi - Museo Storico della Fisica e Centro Studi e Ricerche Enrico Fermi and Istituto Nazionale di Fisica Nucleare (INFN), Italy;
Institute for Innovative Science and Technology , Nagasaki Institute of Applied Science (IIST), Japan Society for the Promotion of Science (JSPS) KAKENHI and Japanese Ministry of Education, Culture, Sports, Science and Technology (MEXT), Japan;
Consejo Nacional de Ciencia (CONACYT) y Tecnolog\'{i}a, through Fondo de Cooperaci\'{o}n Internacional en Ciencia y Tecnolog\'{i}a (FONCICYT) and Direcci\'{o}n General de Asuntos del Personal Academico (DGAPA), Mexico;
Nederlandse Organisatie voor Wetenschappelijk Onderzoek (NWO), Netherlands;
The Research Council of Norway, Norway;
Commission on Science and Technology for Sustainable Development in the South (COMSATS), Pakistan;
Pontificia Universidad Cat\'{o}lica del Per\'{u}, Peru;
Ministry of Science and Higher Education and National Science Centre, Poland;
Korea Institute of Science and Technology Information and National Research Foundation of Korea (NRF), Republic of Korea;
Ministry of Education and Scientific Research, Institute of Atomic Physics and Romanian National Agency for Science, Technology and Innovation, Romania;
Joint Institute for Nuclear Research (JINR), Ministry of Education and Science of the Russian Federation and National Research Centre Kurchatov Institute, Russia;
Ministry of Education, Science, Research and Sport of the Slovak Republic, Slovakia;
National Research Foundation of South Africa, South Africa;
Centro de Aplicaciones Tecnol\'{o}gicas y Desarrollo Nuclear (CEADEN), Cubaenerg\'{\i}a, Cuba, Ministerio de Ciencia e Innovacion and Centro de Investigaciones Energ\'{e}ticas, Medioambientales y Tecnol\'{o}gicas (CIEMAT), Spain;
Swedish Research Council (VR) and Knut \& Alice Wallenberg Foundation (KAW), Sweden;
European Organization for Nuclear Research, Switzerland;
National Science and Technology Development Agency (NSDTA), Suranaree University of Technology (SUT) and Office of the Higher Education Commission under NRU project of Thailand, Thailand;
Turkish Atomic Energy Agency (TAEK), Turkey;
National Academy of  Sciences of Ukraine, Ukraine;
Science and Technology Facilities Council (STFC), United Kingdom;
National Science Foundation of the United States of America (NSF) and United States Department of Energy, Office of Nuclear Physics (DOE NP), United States of America.    
\end{acknowledgement}

\bibliographystyle{utphys}   
\bibliography{bibliography}

\newpage
\appendix
\section{The ALICE Collaboration}
\label{app:collab}



\begingroup
\small
\begin{flushleft}
S.~Acharya$^\textrm{\scriptsize 139}$,
D.~Adamov\'{a}$^\textrm{\scriptsize 96}$,
J.~Adolfsson$^\textrm{\scriptsize 34}$,
M.M.~Aggarwal$^\textrm{\scriptsize 101}$,
G.~Aglieri Rinella$^\textrm{\scriptsize 35}$,
M.~Agnello$^\textrm{\scriptsize 31}$,
N.~Agrawal$^\textrm{\scriptsize 48}$,
Z.~Ahammed$^\textrm{\scriptsize 139}$,
N.~Ahmad$^\textrm{\scriptsize 17}$,
S.U.~Ahn$^\textrm{\scriptsize 80}$
% ...
\renewcommand\labelenumi{\textsuperscript{\theenumi}~}

\section*{Affiliation notes}
\renewcommand\theenumi{\roman{enumi}}
\begin{Authlist}
\item \Adef{0}Deceased
\item \Adef{idp1804176}{Also at: Dipartimento DET del Politecnico di Torino, Turin, Italy}
\item \Adef{idp1823568}{Also at: Georgia State University, Atlanta, Georgia, United States}
\item \Adef{idp4113296}{Also at: M.V. Lomonosov Moscow State University, D.V. Skobeltsyn Institute of Nuclear, Physics, Moscow, Russia}
\item \Adef{idp4471792}{Also at: Department of Applied Physics, Aligarh Muslim University, Aligarh, India}
\item \Adef{idp5449104}{Also at: Institute of Theoretical Physics, University of Wroclaw, Poland}
\end{Authlist}

\section*{Collaboration Institutes}
\renewcommand\theenumi{\arabic{enumi}~}

$^{1}$A.I. Alikhanyan National Science Laboratory (Yerevan Physics Institute) Foundation, Yerevan, Armenia
\\
$^{2}$Benem\'{e}rita Universidad Aut\'{o}noma de Puebla, Puebla, Mexico
\\
$^{3}$Bogolyubov Institute for Theoretical Physics, Kiev, Ukraine
\\
$^{4}$Bose Institute, Department of Physics 
and Centre for Astroparticle Physics and Space Science (CAPSS), Kolkata, India
\\
$^{5}$Budker Institute for Nuclear Physics, Novosibirsk, Russia
\\
$^{6}$California Polytechnic State University, San Luis Obispo, California, United States
\\
$^{7}$Central China Normal University, Wuhan, China
\\
$^{8}$Centre de Calcul de l'IN2P3, Villeurbanne, Lyon, France
\\
$^{9}$Centro de Aplicaciones Tecnol\'{o}gicas y Desarrollo Nuclear (CEADEN), Havana, Cuba
\\
$^{10}$Centro de Investigaciones Energ\'{e}ticas Medioambientales y Tecnol\'{o}gicas (CIEMAT), Madrid, Spain
\\
$^{11}$Centro de Investigaci\'{o}n y de Estudios Avanzados (CINVESTAV), Mexico City and M\'{e}rida, Mexico
\\
$^{12}$Centro Fermi - Museo Storico della Fisica e Centro Studi e Ricerche ``Enrico Fermi', Rome, Italy
\\
$^{13}$Chicago State University, Chicago, Illinois, United States
\\
$^{14}$China Institute of Atomic Energy, Beijing, China
\\
$^{15}$COMSATS Institute of Information Technology (CIIT), Islamabad, Pakistan
\\
$^{16}$Departamento de F\'{\i}sica de Part\'{\i}culas and IGFAE, Universidad de Santiago de Compostela, Santiago de Compostela, Spain
\\
$^{17}$Department of Physics, Aligarh Muslim University, Aligarh, India
\\
$^{18}$Department of Physics, Ohio State University, Columbus, Ohio, United States
\\
$^{19}$Department of Physics, Pusan National University, Pusan, South Korea
\\
$^{20}$Department of Physics, Sejong University, Seoul, South Korea
\\
$^{21}$Department of Physics, University of Oslo, Oslo, Norway
\\
$^{22}$Department of Physics and Technology, University of Bergen, Bergen, Norway
\\
$^{23}$Dipartimento di Fisica dell'Universit\`{a} 'La Sapienza'
and Sezione INFN, Rome, Italy
\\
$^{24}$Dipartimento di Fisica dell'Universit\`{a}
and Sezione INFN, Cagliari, Italy
\\
$^{25}$Dipartimento di Fisica dell'Universit\`{a}
and Sezione INFN, Trieste, Italy
\\
$^{26}$Dipartimento di Fisica dell'Universit\`{a}
and Sezione INFN, Turin, Italy
\\
$^{27}$Dipartimento di Fisica e Astronomia dell'Universit\`{a}
and Sezione INFN, Bologna, Italy
\\
$^{28}$Dipartimento di Fisica e Astronomia dell'Universit\`{a}
and Sezione INFN, Catania, Italy
\\
$^{29}$Dipartimento di Fisica e Astronomia dell'Universit\`{a}
and Sezione INFN, Padova, Italy
\\
$^{30}$Dipartimento di Fisica `E.R.~Caianiello' dell'Universit\`{a}
and Gruppo Collegato INFN, Salerno, Italy
\\
$^{31}$Dipartimento DISAT del Politecnico and Sezione INFN, Turin, Italy
\\
$^{32}$Dipartimento di Scienze e Innovazione Tecnologica dell'Universit\`{a} del Piemonte Orientale and INFN Sezione di Torino, Alessandria, Italy
\\
$^{33}$Dipartimento Interateneo di Fisica `M.~Merlin'
and Sezione INFN, Bari, Italy
\\
$^{34}$Division of Experimental High Energy Physics, University of Lund, Lund, Sweden
\\
$^{35}$European Organization for Nuclear Research (CERN), Geneva, Switzerland
\\
$^{36}$Excellence Cluster Universe, Technische Universit\"{a}t M\"{u}nchen, Munich, Germany
\\
$^{37}$Faculty of Engineering, Bergen University College, Bergen, Norway
\\
$^{38}$Faculty of Mathematics, Physics and Informatics, Comenius University, Bratislava, Slovakia
\\
$^{39}$Faculty of Nuclear Sciences and Physical Engineering, Czech Technical University in Prague, Prague, Czech Republic
\\
$^{40}$Faculty of Science, P.J.~\v{S}af\'{a}rik University, Ko\v{s}ice, Slovakia
\\
$^{41}$Faculty of Technology, Buskerud and Vestfold University College, Tonsberg, Norway
\\
$^{42}$Frankfurt Institute for Advanced Studies, Johann Wolfgang Goethe-Universit\"{a}t Frankfurt, Frankfurt, Germany
\\
$^{43}$Gangneung-Wonju National University, Gangneung, South Korea
\\
$^{44}$Gauhati University, Department of Physics, Guwahati, India
\\
$^{45}$Helmholtz-Institut f\"{u}r Strahlen- und Kernphysik, Rheinische Friedrich-Wilhelms-Universit\"{a}t Bonn, Bonn, Germany
\\
$^{46}$Helsinki Institute of Physics (HIP), Helsinki, Finland
\\
$^{47}$Hiroshima University, Hiroshima, Japan
\\
$^{48}$Indian Institute of Technology Bombay (IIT), Mumbai, India
\\
$^{49}$Indian Institute of Technology Indore, Indore, India
\\
$^{50}$Indonesian Institute of Sciences, Jakarta, Indonesia
\\
$^{51}$INFN, Laboratori Nazionali di Frascati, Frascati, Italy
\\
$^{52}$INFN, Laboratori Nazionali di Legnaro, Legnaro, Italy
\\
$^{53}$INFN, Sezione di Bari, Bari, Italy
\\
$^{54}$INFN, Sezione di Bologna, Bologna, Italy
\\
$^{55}$INFN, Sezione di Cagliari, Cagliari, Italy
\\
$^{56}$INFN, Sezione di Catania, Catania, Italy
\\
$^{57}$INFN, Sezione di Padova, Padova, Italy
\\
$^{58}$INFN, Sezione di Roma, Rome, Italy
\\
$^{59}$INFN, Sezione di Torino, Turin, Italy
\\
$^{60}$INFN, Sezione di Trieste, Trieste, Italy
\\
$^{61}$Inha University, Incheon, South Korea
\\
$^{62}$Institut de Physique Nucl\'eaire d'Orsay (IPNO), Universit\'e Paris-Sud, CNRS-IN2P3, Orsay, France
\\
$^{63}$Institute for Nuclear Research, Academy of Sciences, Moscow, Russia
\\
$^{64}$Institute for Subatomic Physics of Utrecht University, Utrecht, Netherlands
\\
$^{65}$Institute for Theoretical and Experimental Physics, Moscow, Russia
\\
$^{66}$Institute of Experimental Physics, Slovak Academy of Sciences, Ko\v{s}ice, Slovakia
\\
$^{67}$Institute of Physics, Academy of Sciences of the Czech Republic, Prague, Czech Republic
\\
$^{68}$Institute of Physics, Bhubaneswar, India
\\
$^{69}$Institute of Space Science (ISS), Bucharest, Romania
\\
$^{70}$Institut f\"{u}r Informatik, Johann Wolfgang Goethe-Universit\"{a}t Frankfurt, Frankfurt, Germany
\\
$^{71}$Institut f\"{u}r Kernphysik, Johann Wolfgang Goethe-Universit\"{a}t Frankfurt, Frankfurt, Germany
\\
$^{72}$Institut f\"{u}r Kernphysik, Westf\"{a}lische Wilhelms-Universit\"{a}t M\"{u}nster, M\"{u}nster, Germany
\\
$^{73}$Instituto de Ciencias Nucleares, Universidad Nacional Aut\'{o}noma de M\'{e}xico, Mexico City, Mexico
\\
$^{74}$Instituto de F\'{i}sica, Universidade Federal do Rio Grande do Sul (UFRGS), Porto Alegre, Brazil
\\
$^{75}$Instituto de F\'{\i}sica, Universidad Nacional Aut\'{o}noma de M\'{e}xico, Mexico City, Mexico
\\
$^{76}$IRFU, CEA, Universit\'{e} Paris-Saclay, Saclay, France
\\
$^{77}$iThemba LABS, National Research Foundation, Somerset West, South Africa
\\
$^{78}$Joint Institute for Nuclear Research (JINR), Dubna, Russia
\\
$^{79}$Konkuk University, Seoul, South Korea
\\
$^{80}$Korea Institute of Science and Technology Information, Daejeon, South Korea
\\
$^{81}$KTO Karatay University, Konya, Turkey
\\
$^{82}$Laboratoire de Physique Corpusculaire (LPC), Clermont Universit\'{e}, Universit\'{e} Blaise Pascal, CNRS--IN2P3, Clermont-Ferrand, France
\\
$^{83}$Laboratoire de Physique Subatomique et de Cosmologie, Universit\'{e} Grenoble-Alpes, CNRS-IN2P3, Grenoble, France
\\
$^{84}$Lawrence Berkeley National Laboratory, Berkeley, California, United States
\\
$^{85}$Moscow Engineering Physics Institute, Moscow, Russia
\\
$^{86}$Nagasaki Institute of Applied Science, Nagasaki, Japan
\\
$^{87}$National and Kapodistrian University of Athens, Physics Department, Athens, Greece, Athens, Greece
\\
$^{88}$National Centre for Nuclear Studies, Warsaw, Poland
\\
$^{89}$National Institute for Physics and Nuclear Engineering, Bucharest, Romania
\\
$^{90}$National Institute of Science Education and Research, Bhubaneswar, India
\\
$^{91}$National Nuclear Research Center, Baku, Azerbaijan
\\
$^{92}$National Research Centre Kurchatov Institute, Moscow, Russia
\\
$^{93}$Niels Bohr Institute, University of Copenhagen, Copenhagen, Denmark
\\
$^{94}$Nikhef, Nationaal instituut voor subatomaire fysica, Amsterdam, Netherlands
\\
$^{95}$Nuclear Physics Group, STFC Daresbury Laboratory, Daresbury, United Kingdom
\\
$^{96}$Nuclear Physics Institute, Academy of Sciences of the Czech Republic, \v{R}e\v{z} u Prahy, Czech Republic
\\
$^{97}$Oak Ridge National Laboratory, Oak Ridge, Tennessee, United States
\\
$^{98}$Petersburg Nuclear Physics Institute, Gatchina, Russia
\\
$^{99}$Physics Department, Creighton University, Omaha, Nebraska, United States
\\
$^{100}$Physics department, Faculty of science, University of Zagreb, Zagreb, Croatia
\\
$^{101}$Physics Department, Panjab University, Chandigarh, India
\\
$^{102}$Physics Department, University of Cape Town, Cape Town, South Africa
\\
$^{103}$Physics Department, University of Jammu, Jammu, India
\\
$^{104}$Physics Department, University of Rajasthan, Jaipur, India
\\
$^{105}$Physikalisches Institut, Eberhard Karls Universit\"{a}t T\"{u}bingen, T\"{u}bingen, Germany
\\
$^{106}$Physikalisches Institut, Ruprecht-Karls-Universit\"{a}t Heidelberg, Heidelberg, Germany
\\
$^{107}$Physik Department, Technische Universit\"{a}t M\"{u}nchen, Munich, Germany
\\
$^{108}$Purdue University, West Lafayette, Indiana, United States
\\
$^{109}$Research Division and ExtreMe Matter Institute EMMI, GSI Helmholtzzentrum f\"ur Schwerionenforschung GmbH, Darmstadt, Germany
\\
$^{110}$Rudjer Bo\v{s}kovi\'{c} Institute, Zagreb, Croatia
\\
$^{111}$Russian Federal Nuclear Center (VNIIEF), Sarov, Russia
\\
$^{112}$Saha Institute of Nuclear Physics, Kolkata, India
\\
$^{113}$School of Physics and Astronomy, University of Birmingham, Birmingham, United Kingdom
\\
$^{114}$Secci\'{o}n F\'{\i}sica, Departamento de Ciencias, Pontificia Universidad Cat\'{o}lica del Per\'{u}, Lima, Peru
\\
$^{115}$SSC IHEP of NRC Kurchatov institute, Protvino, Russia
\\
$^{116}$Stefan Meyer Institut f\"{u}r Subatomare Physik (SMI), Vienna, Austria
\\
$^{117}$SUBATECH, IMT Atlantique, Universit\'{e} de Nantes, CNRS-IN2P3, Nantes, France
\\
$^{118}$Suranaree University of Technology, Nakhon Ratchasima, Thailand
\\
$^{119}$Technical University of Ko\v{s}ice, Ko\v{s}ice, Slovakia
\\
$^{120}$Technical University of Split FESB, Split, Croatia
\\
$^{121}$The Henryk Niewodniczanski Institute of Nuclear Physics, Polish Academy of Sciences, Cracow, Poland
\\
$^{122}$The University of Texas at Austin, Physics Department, Austin, Texas, United States
\\
$^{123}$Universidad Aut\'{o}noma de Sinaloa, Culiac\'{a}n, Mexico
\\
$^{124}$Universidade de S\~{a}o Paulo (USP), S\~{a}o Paulo, Brazil
\\
$^{125}$Universidade Estadual de Campinas (UNICAMP), Campinas, Brazil
\\
$^{126}$Universidade Federal do ABC, Santo Andre, Brazil
\\
$^{127}$University of Houston, Houston, Texas, United States
\\
$^{128}$University of Jyv\"{a}skyl\"{a}, Jyv\"{a}skyl\"{a}, Finland
\\
$^{129}$University of Liverpool, Liverpool, United Kingdom
\\
$^{130}$University of Tennessee, Knoxville, Tennessee, United States
\\
$^{131}$University of the Witwatersrand, Johannesburg, South Africa
\\
$^{132}$University of Tokyo, Tokyo, Japan
\\
$^{133}$University of Tsukuba, Tsukuba, Japan
\\
$^{134}$Universit\'{e} de Lyon, Universit\'{e} Lyon 1, CNRS/IN2P3, IPN-Lyon, Villeurbanne, Lyon, France
\\
$^{135}$Universit\'{e} de Strasbourg, CNRS, IPHC UMR 7178, F-67000 Strasbourg, France, Strasbourg, France
\\
$^{136}$Universit\`{a} degli Studi di Pavia, Pavia, Italy
\\
$^{137}$Universit\`{a} di Brescia, Brescia, Italy
\\
$^{138}$V.~Fock Institute for Physics, St. Petersburg State University, St. Petersburg, Russia
\\
$^{139}$Variable Energy Cyclotron Centre, Kolkata, India
\\
$^{140}$Warsaw University of Technology, Warsaw, Poland
\\
$^{141}$Wayne State University, Detroit, Michigan, United States
\\
$^{142}$Wigner Research Centre for Physics, Hungarian Academy of Sciences, Budapest, Hungary
\\
$^{143}$Yale University, New Haven, Connecticut, United States
\\
$^{144}$Yonsei University, Seoul, South Korea
\\
$^{145}$Zentrum f\"{u}r Technologietransfer und Telekommunikation (ZTT), Fachhochschule Worms, Worms, Germany
\endgroup

\end{document}